\documentclass[aps,
  prl,
  showpacs,
  superscriptaddress,
  floatfix,
  twocolumn,
  secnumarabic,
  amssymb,
  amsmath]{revtex4}

\pdfoutput=1

\usepackage{ifpdf}
\usepackage{graphicx}
\usepackage{dcolumn}
\usepackage{amsmath}
\usepackage{amsfonts}
\usepackage{amssymb}
\usepackage{bm}
\usepackage{color}
\usepackage{ulem}
\usepackage{longtable}
\usepackage{sidecap}

\begin{document}

\preprint{A. Nielsen {\it et al.}, version: 12 August 2008}

\title{All Oxide Ferromagnet/Semiconductor Epitaxial Heterostructures}

\author{A. Nielsen}
 \affiliation{Walther-Mei{\ss}ner-Institut, Bayerische Akademie der
              Wissenschaften, D-85748 Garching, Germany}

\author{A. Brandlmaier}
 \affiliation{Walther-Mei{\ss}ner-Institut, Bayerische Akademie der
              Wissenschaften, D-85748 Garching, Germany}

\author{M. Althammer}
 \affiliation{Walther-Mei{\ss}ner-Institut, Bayerische Akademie der
              Wissenschaften, D-85748 Garching, Germany}

\author{W. Kaiser}
 \affiliation{Walther-Mei{\ss}ner-Institut, Bayerische Akademie der
              Wissenschaften, D-85748 Garching, Germany}

\author{M. Opel}
 \affiliation{Walther-Mei{\ss}ner-Institut, Bayerische Akademie der
              Wissenschaften, D-85748 Garching, Germany}

\author{J. Simon}
\affiliation{Institut f\"{u}r Anorganische Chemie, Universit\"{a}t Bonn, D-53117 Bonn,
Germany}

\author{W. Mader}
\affiliation{Institut f\"{u}r Anorganische Chemie, Universit\"{a}t Bonn, D-53117 Bonn,
Germany}

\author{S.T.B. Goennenwein}
 \affiliation{Walther-Mei{\ss}ner-Institut, Bayerische Akademie der Wissenschaften,
              D-85748 Garching, Germany}
 \affiliation{Physik-Department, Technische Universit\"{a}t M\"{u}nchen, D-85748
              Garching, Germany}

\author{R. Gross}
 \email{Rudolf.Gross@wmi.badw.de}
 \affiliation{Walther-Mei{\ss}ner-Institut, Bayerische Akademie der Wissenschaften,
              D-85748 Garching, Germany}
 \affiliation{Physik-Department, Technische Universit\"{a}t M\"{u}nchen, D-85748
              Garching, Germany}

\date{12 August 2008}

\begin{abstract}
Oxide based ferromagnet/semiconductor heterostructures offer substantial
advantages for spin electronics. We have grown (111) oriented Fe$_3$O$_4$ thin
films and Fe$_3$O$_4$/ZnO heterostructures on ZnO(0001) and Al$_2$O$_3$(0001)
substrates by pulsed laser deposition. High quality crystalline films with
mosaic spread as small as 0.03$^{\circ}$, sharp interfaces, and rms surface
roughness of 0.3\,nm were achieved. Magnetization measurements show clear
ferromagnetic behavior of the magnetite layers with a saturation magnetization
of 3.2\,$\mu_{\rm{B}}$/f.u. at 300\,K. Our results demonstrate that the
$\mathrm{Fe}_{3}\mathrm{O}_{4}/\mathrm{ZnO}$ system is an intriguing and
promising candidate for the realization of multi-functional heterostructures.
\end{abstract}


 \maketitle

In spin electronics the spin degree of freedom is exploited to realize
electronic devices with novel or superior functionality
\cite{Wolf:2001a,Zutic:2004a}. For {\it semiconductor} spintronic devices,
charge carrier populations with a controllable spin polarization must be
created within conventional semiconductors. A seemingly straightforward
approach to do so is to inject spin polarized carriers from a ferromagnetic
electrode into the semiconductor material. Unfortunately, the large
conductivity mismatch between conventional metallic $3d$ ferromagnets and
semiconductors prevents an efficient spin injection \cite{Schmidt:2000}. This
problem can be circumvented e.g.~via the introduction of Schottky or tunnel
barriers at the ferromagnet/semiconductor (FM/SC) interface
\cite{Hanbicki:2003}. An alternative approach is to use ferromagnetic materials
with very high spin polarization and small conductivity mismatch with
semiconductors. Ferromagnets with a spin polarization of $100\%$ -- so-called
half metals -- thus are most interesting. The oxide ferrimagnet Fe$_3$O$_4$,
onto which we focus here, is a half metal according to band structure
calculations \cite{Zhang:1991}, and a spin polarization of up to
-(80\,$\pm$\,5)\,\% has been reported from spin-resolved photoelectron
spectroscopy experiments in $(111)$-oriented $\mathrm{Fe}_{3}\mathrm{O}_{4}$
\cite{Dedkov:2002,Fonin:2005}. Furthermore, the conductivity $\sigma\approx
200\,\Omega^{-1}\mathrm{cm}^{-1}$ of $\mathrm{Fe}_{3}\mathrm{O}_{4}$ at room
temperature is low \cite{Reisinger:2004a}, while the Curie temperature
$T_{\mathrm{C}}\simeq 860\,\mathrm{K}$ is well above room temperature, making
$\mathrm{Fe}_{3}\mathrm{O}_{4}$ a promising material for spin injection into
semiconductors. However, the $\mathrm{Fe}_{3}\mathrm{O}_{4}$/semiconductor
heterostructures investigated so far
\cite{Kennedy:1999,Reisinger:2003b,Lu:2004,Lu:2005,Watts:2004,Boothman:2007}
show that for both group IV and III-V semiconductors, it is very difficult to
grow $\mathrm{Fe}_{3}\mathrm{O}_{4}$ thin films with high crystalline quality,
while preventing the formation of secondary phases at the FM/SC interface. To
our knowledge, the deposition of $\mathrm{Fe}_{3}\mathrm{O}_{4}$ onto a II-VI
semiconductor has not been reported to date.

In this letter, we show that $(111)$-oriented $\mathrm{Fe}_{3}\mathrm{O}_{4}$
can be epitaxially grown onto the II-VI semiconductor ZnO using pulsed laser
deposition (PLD). Furthermore, we demonstrate the epitaxial growth of ZnO thin
films onto $(111)$-oriented $\mathrm{Fe}_{3}\mathrm{O}_{4}$. The FM/SC
heterostructures thus obtained are characterized using x-ray diffraction (XRD),
atomic force microscopy (AFM), transmission electron microscopy (TEM), and
superconducting quantum interference device (SQUID) magnetometry. Our results
are compared to the properties of $(111)$-oriented
$\mathrm{Fe}_{3}\mathrm{O}_{4}$ films on $\mathrm{Al}_{2}\mathrm{O}_{3}$
substrates, which have been extensively investigated and can be considered as a
benchmark system
\cite{Dellile:1996,Margulies:1996,Ogale:1998,Yamaguchi:2002,Farrow:2003,Moussy:2004}.
We find that the magnetic and structural properties of our $(111)$
$\mathrm{Fe}_{3}\mathrm{O}_{4}$ films on ZnO are state of the art, with sharp
FM/SC interfaces. This makes them promising for spin injection devices.

\begin{figure}
\includegraphics[width=0.8\columnwidth]{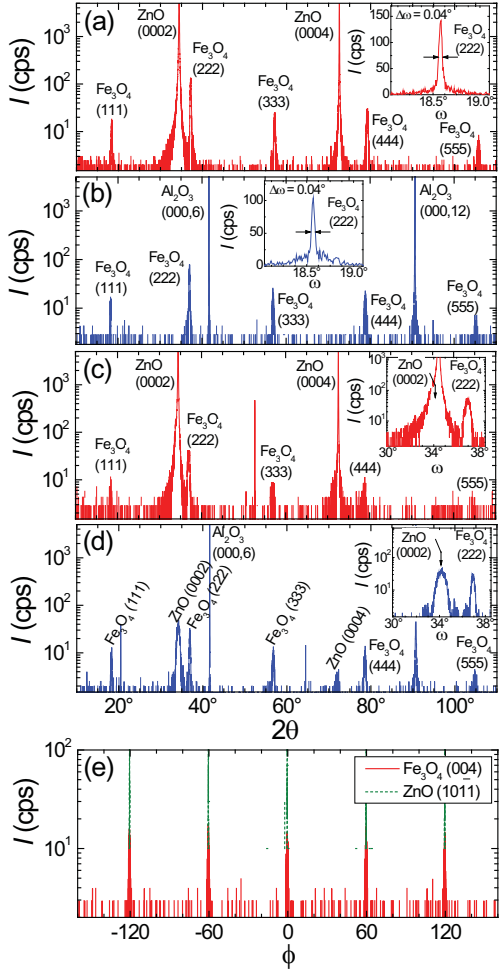} 
\caption{\label{fig:xraymitphi}
        $\omega$-$2\theta$ scans of a 31\,nm thick Fe$_3$O$_4$(111) film
        on ZnO$(0001)$ (a), a 34\,nm thick Fe$_3$O$_4$(111) film
        on $\mathrm{Al}_{2}\mathrm{O}_{3}$$(0001)$ (b), a ZnO/Fe$_3$O$_4$//ZnO (29\,nm/18\,nm) heterostructure (c),
        and a ZnO/Fe$_3$O$_4$//Al$_2$O$_3$ (11\,nm/28\,nm) heterostructure (d).
        In (e) $\phi$ scans of the Fe$_3$O$_4$ (004) and the ZnO
        (10$\overline{1}$1) reflections of the Fe$_3$O$_4$ film on ZnO (cf.~panel (b)) are shown.
        }
\end{figure}

We have grown four different types of epitaxial Fe$_3$O$_4$-based thin film
samples by pulsed laser deposition \cite{Klein:1999b}: (i) single Fe$_3$O$_4$
layers, deposited on ZnO$(0001)$ substrates, (ii) single Fe$_3$O$_4$ layers on
Al$_2$O$_3$$(0001)$ substrates, (iii)
$\mathrm{ZnO}/\mathrm{Fe}_{3}\mathrm{O}_{4}$ thin film heterostructures on
ZnO$(0001)$, and (iv) $\mathrm{ZnO}/\mathrm{Fe}_{3}\mathrm{O}_{4}$
heterostructures on Al$_2$O$_3$$(0001)$. In the following, we will refer to
these samples as $\mathrm{Fe}_{3}\mathrm{O}_{4}//\mathrm{ZnO}$,
$\mathrm{Fe}_{3}\mathrm{O}_{4}//\mathrm{Al}_{2}\mathrm{O}_{3}$,
$\mathrm{ZnO}/\mathrm{Fe}_{3}\mathrm{O}_{4}//\mathrm{ZnO}$, and
$\mathrm{ZnO}/\mathrm{Fe}_{3}\mathrm{O}_{4}//\mathrm{Al}_{2}\mathrm{O}_{3}$,
respectively. Both the Fe$_3$O$_4$ and the ZnO layers were deposited from
stoichiometric Fe$_3$O$_4$ and ZnO targets, respectively, using identical
growth parameters. All samples were grown in pure Ar atmosphere at a pressure
of  $3.7\times 10^{-3}\,\mathrm{mbar}$ and a substrate temperature of 590\,K.
The energy density of the KrF excimer laser pulses (wavelength
$\lambda=248$\,nm) on the respective targets was 2\,J/cm$^2$ at a pulse
repetition rate of 2\,Hz.

We first address the  structural and magnetic properties of the
$\mathrm{Fe}_{3}\mathrm{O}_{4}//\mathrm{ZnO}$ and
$\mathrm{Fe}_{3}\mathrm{O}_{4}//\mathrm{Al}_{2}\mathrm{O}_{3}$ single layer
films. Figure~\ref{fig:xraymitphi} shows representative $\omega$-$2\theta$
scans of all four sample types. Set aside the substrate reflections, only
$\mathrm{Fe}_{3}\mathrm{O}_{4}$$(\ell\ell\ell)$ and ZnO$(000\ell)$ reflections
are observed. This shows that magnetite grows $(111)$-oriented on ZnO $(0001)$
substrates, in close analogy to films on $\mathrm{Al}_{2}\mathrm{O}_{3}$
$(0001)$ substrates \cite{Yamaguchi:2002}. Furthermore, in the ZnO/Fe$_3$O$_4$
heterostructures the ZnO layer grows $(0001)$ oriented on the $(111)$ oriented
$\mathrm{Fe}_{3}\mathrm{O}_{4}$ layer. The fact that
$\mathrm{Fe}_{3}\mathrm{O}_{4}$ grows on both substrates in $(111)$ orientation
is astonishing. Bulk Fe$_3$O$_4$ showing the inverse spinel structure ($Fd3m$)
is cubic with lattice constant of $a_{{\rm Fe}_3{\rm O}_4}=8.3963$\,{\AA}
\cite{Fleet:1981}. ZnO has hexagonal wurzite structure with $a_{\rm
ZnO}=3.2498$\,{\AA} \cite{Reeber:1970} and sapphire hexagonal structure
($R\overline{3}c$) with $a_{\rm Al_2O_3}=4.759$\,{\AA} in the hexagonal cell
\cite{Kirfel:1990}. Therefore, there is a large lattice mismatch between
multiples of the $a$-axis lattice parameters of both substrates and $\sqrt{2}a$
for the $(111)$ oriented Fe$_3$O$_4$. The same is true for other Fe$_3$O$_4$
orientations. Nevertheless, $\mathrm{Fe}_{3}\mathrm{O}_{4}$ tends to grow in
$(111)$ orientation with well defined epitaxial relations on substrates which
are not well lattice matched \cite{Kennedy:1999,Watts:2004}. To quantify the
Fe$_3$O$_4$ lattice parameters, the $(111)$ lattice plane distance $d_{111}$
was evaluated. For $\mathrm{Fe}_{3}\mathrm{O}_{4}//\mathrm{ZnO}$ samples, this
leads to $d_{111}=\left(4.836\pm 0.005\right)\mathrm{{\AA}}$, and for
$\mathrm{Fe}_{3}\mathrm{O}_{4}//\mathrm{Al}_{2}\mathrm{O}_{3}$ we obtain
$d_{111}=\left(4.852\pm 0.005\right)\mathrm{{\AA}}$. These values are close to
$d_{111,\mathrm{bulk}}=4.848\mathrm{{\AA}}$ of bulk $\mathrm{Fe}_{3}\mathrm{O}_{4}$
\cite{Fleet:1981}, suggesting that the films grow fully relaxed. Small
deviations of the film $d_{111}$ from $d_{111,\mathrm{bulk}}$ have been
reported for growth on Si, GaAs, $\mathrm{Al}_{2}\mathrm{O}_{3}$, and are
attributed to substrate-induced strain. The in-plane epitaxial relations
between the $\mathrm{Fe}_{3}\mathrm{O}_{4}$ film and the substrates have been
derived from $\phi$-scans. For $\mathrm{Fe}_{3}\mathrm{O}_{4}//\mathrm{ZnO}$,
the $\{004\}$ reflections of magnetite appears every $60^{\circ}$ at the same
$\phi$ angles as the ZnO $\{10\overline{1}1\}$ reflection
(Fig.~\ref{fig:xraymitphi}(e)). This demonstrates the good in-plane orientation
of $\mathrm{Fe}_{3}\mathrm{O}_{4}$$(111)$ films on ZnO, with the ZnO
$[2\overline{1}\overline{1}0]$ close-packed direction parallel to the
Fe$_3$O$_4$ $[1\overline{1}0]$ close-packed direction. The same parallelism of
the close-packed oxygen sub-lattices is obtained for the epitaxial
$\mathrm{Fe}_{3}\mathrm{O}_{4}//\mathrm{Al}_{2}\mathrm{O}_{3}$ film. The good
structural quality of the samples is further evidenced by the small full width
at half maximum (FWHM) of the $(222)$ (out of plane) rocking curve
$\Delta\omega = 0.04^\circ$ in both
$\mathrm{Fe}_{3}\mathrm{O}_{4}//\mathrm{ZnO}$ and
$\mathrm{Fe}_{3}\mathrm{O}_{4}//\mathrm{Al}_{2}\mathrm{O}_{3}$. Within the film
plane, the mosaicity is somewhat larger, with $\Delta \omega \simeq 0.9^\circ$
for the $(004)$ $\mathrm{Fe}_{3}\mathrm{O}_{4}$ reflection in
$\mathrm{Fe}_{3}\mathrm{O}_{4}//\mathrm{ZnO}$. The surface roughness of the
films was studied using atomic force microscopy at room temperature. For the
$\mathrm{Fe}_{3}\mathrm{O}_{4}//\mathrm{ZnO}$ sample, a rms roughness of
0.3\,nm is found, with a peak to peak value of 0.8\,nm in an area of $0.9
\times 0.9\,\mu\textrm{m}^2$ (not shown). This low surface roughness is an
important prerequisite for the realization of
$\mathrm{Fe}_{3}\mathrm{O}_{4}$/ZnO FM/SC heterostructures with smooth
interfaces.

\begin{figure}[b]
\includegraphics[width=0.8\columnwidth]{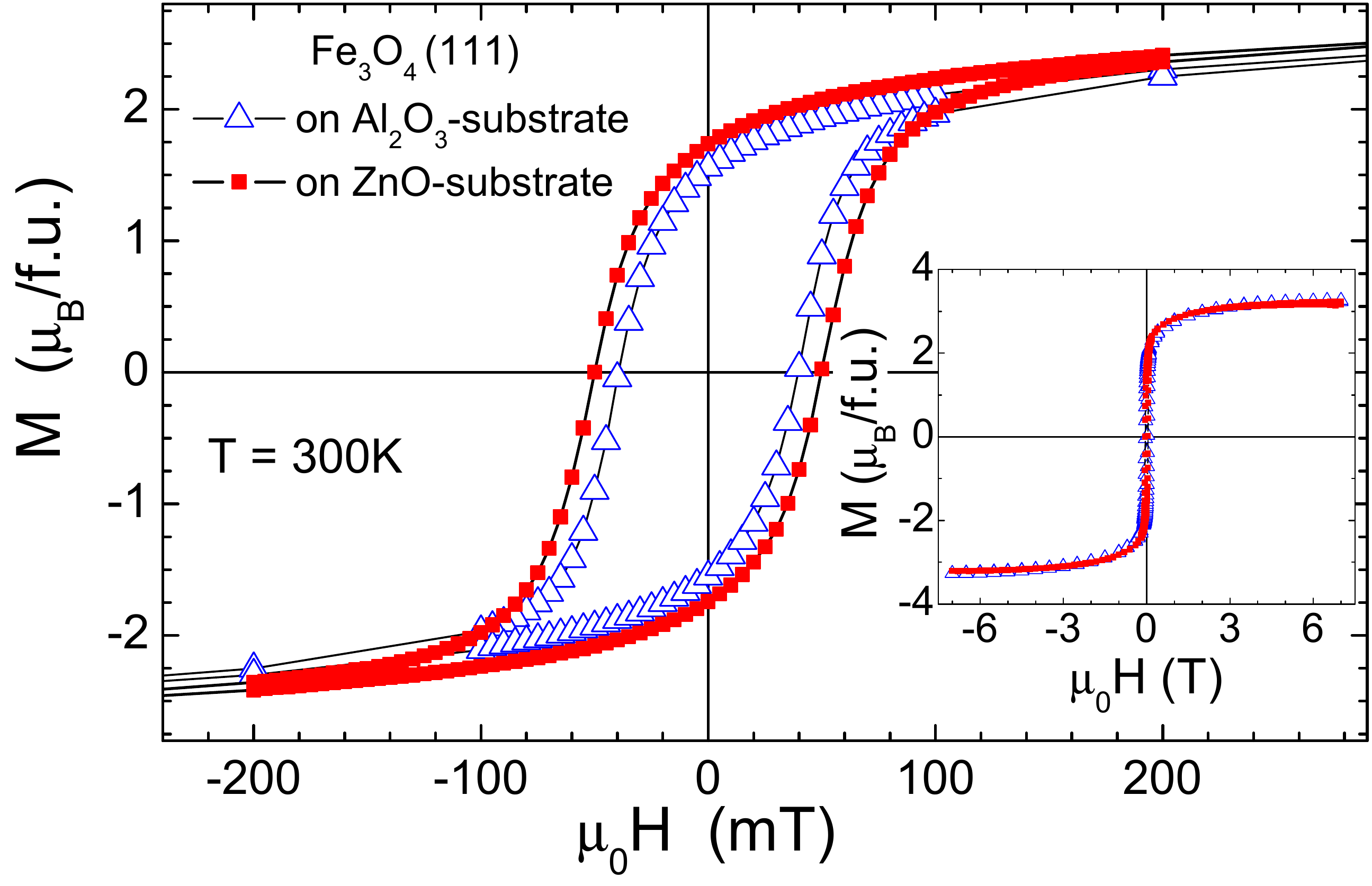}
   \caption{
   $M(H)$ curves of Fe$_3$O$_4$ films grown on (a) Al$_2$O$_3$
   (open triangles, 34\,nm thick) and (b) ZnO (full squares, 31\,nm thick), taken
   at 300\,K with $H \| [1\overline{1}0]$ of Fe$_3$O$_4$.
   The inset shows the same data on an enlarged field scale.}
   \label{fig:MH}
\end{figure}

The magnetic properties of the Fe$_3$O$_4$ thin films were investigated using a
SQUID magnetometer. Figure~\ref{fig:MH} shows the magnetization $M$ versus
magnetic field $H$ of the $\mathrm{Fe}_{3}\mathrm{O}_{4}//\mathrm{ZnO}$ and
$\mathrm{Fe}_{3}\mathrm{O}_{4}//\mathrm{Al}_{2}\mathrm{O}_{3}$ samples
discussed above (cf.~Fig.~\ref{fig:xraymitphi}). The diamagnetic contribution
of the substrates has been subtracted. The magnetic hysteresis loops of both
films show clear ferromagnetic hysteresis. The coercivity
$\mu_{0}H_{\mathrm{c}}=50\,\mathrm{mT}$ and the remanence $M_{\mathrm{R}}=
1.74\,\mu_{\mathrm{B}}/\mathrm{f.u.}$ for
$\mathrm{Fe}_{3}\mathrm{O}_{4}//\mathrm{ZnO}$ are closely comparable to
$\mu_{0}H_{\mathrm{c}}= 40\,\mathrm{mT}$ and $M_{\mathrm{R}}=
1.55\,\mu_{\mathrm{B}}/\mathrm{f.u.}$ for
$\mathrm{Fe}_{3}\mathrm{O}_{4}//\mathrm{Al}_{2}\mathrm{O}_{3}$. Note that these
coercive fields are well within the range $30\,\mathrm{mT} \le
\mu_{0}H_{\mathrm{c}} \le 60\,\mathrm{mT}$ reported for $(111)$ oriented
Fe$_3$O$_4$ films
grown on sapphire \cite{Dellile:1996,Yamaguchi:2002,Farrow:2003,Moussy:2004}.
At $\mu_{0}H=7\,\mathrm{T}$, the magnetization reaches
$3.2\,\mu_{\mathrm{B}}/\mathrm{f.u.}$ in both films (Fig.~\ref{fig:MH}, inset),
but does not fully saturate. This most likely originates from the presence of
antiphase boundaries \cite{Margulies:1996,Moussy:2004}. Taken together, the
magnetometry results show that the $(111)$-oriented
$\mathrm{Fe}_{3}\mathrm{O}_{4}$ thin films grown on ZnO can be considered as
state of the art also in terms of their magnetic properties.

\begin{figure}[tb]
\includegraphics[width=0.8\columnwidth]{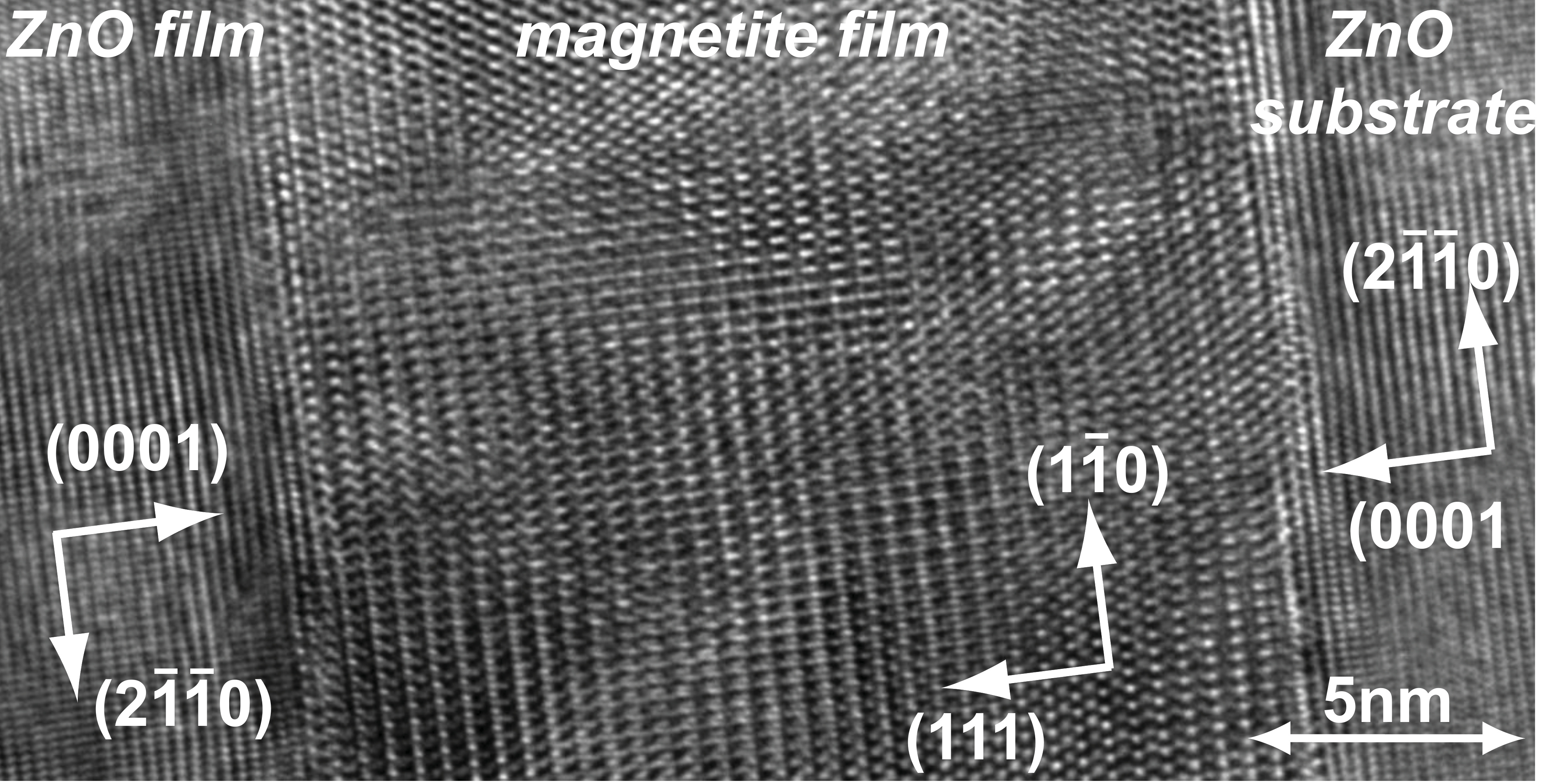}
    \caption{
    HRTEM image of an 18\,nm thick (111) oriented Fe$_3$O$_4$ film on a (0001) ZnO substrate
    (right) with a 29\,nm thick (000$\overline{1}$) oriented ZnO film on top (left).}
    \label{fig:TEM}
\end{figure}

Having established that $(111)$ oriented $\mathrm{Fe}_{3}\mathrm{O}_{4}$ thin
films with excellent structural, magnetic, and surface properties can be grown
on ZnO$(0001)$  substrates, we further demonstrate that it is also possible to
grow ZnO films epitaxially onto $(111)$-oriented
$\mathrm{Fe}_{3}\mathrm{O}_{4}$. In this way fully epitaxial
$\mathrm{Fe}_{3}\mathrm{O}_{4}/\mathrm{ZnO}$ heterostructures can be realized.
Figure~\ref{fig:xraymitphi}(c) and (d) show $\omega$-$2\theta$ x-ray
diffraction scans of representative
$\mathrm{ZnO}/\mathrm{Fe}_{3}\mathrm{O}_{4}//\mathrm{ZnO}$ and
$\mathrm{ZnO}/\mathrm{Fe}_{3}\mathrm{O}_{4}//\mathrm{Al}_{2}\mathrm{O}_{3}$
samples. For the
$\mathrm{ZnO}/\mathrm{Fe}_{3}\mathrm{O}_{4}//\mathrm{Al}_{2}\mathrm{O}_{3}$
sample, the $(000\ell)$ ZnO film reflections are not masked by the substrate
reflections and can thus be unambiguously identified. Instead of going through
the details of the structural and magnetic characterization once more, let us
note that the structural quality of the heterostructure samples again is very
good, matching the crystalline properties of the single
$\mathrm{Fe}_{3}\mathrm{O}_{4}$ layers discussed above. In the following we
will discuss the results of the TEM analysis. Figure~\ref{fig:TEM} shows a TEM
cross-sectional micrograph of the
$\mathrm{ZnO}/\mathrm{Fe}_{3}\mathrm{O}_{4}//\mathrm{ZnO}$ sample. Electron
diffraction shows that the ZnO film is ($000\overline{1}$) oriented, whereas
the ZnO substrate is ($0001$) oriented. Hence, the $c$ axes of the polar ZnO
crystals of both substrate and film point towards the magnetite film which is
important for a stable interface and which can be explained by arguments of
structural chemistry. Smooth and abrupt interfaces between the ZnO and
$\mathrm{Fe}_{3}\mathrm{O}_{4}$ layers are clearly evident from
Fig.~\ref{fig:TEM}, corroborating the conclusions drawn from the XRD analysis.

In summary, we have grown Fe$_3$O$_4$(111) films and fully epitaxial
$\mathrm{ZnO} (0001)/\mathrm{Fe}_{3}\mathrm{O}_{4} (111)$ heterostructures on
$(0001)$-oriented ZnO and $\mathrm{Al}_{2}\mathrm{O}_{3}$ substrates. While the
deposition of $\mathrm{Fe}_{3}\mathrm{O}_{4}$ on
$\mathrm{Al}_{2}\mathrm{O}_{3}$ is established, the growth of
$\mathrm{Fe}_{3}\mathrm{O}_{4}/\mathrm{ZnO}$ FM/SC heterostructures has not yet
been explored to our knowledge. Combining XRD, AFM, TEM, and SQUID
magnetometry, we show that the Fe$_3$O$_4$ films are state of the art in terms
of their structural and magnetic properties, with smooth and abrupt interfaces
between $\mathrm{Fe}_{3}\mathrm{O}_{4}$ and ZnO. This suggests that the
$\mathrm{Fe}_{3}\mathrm{O}_{4}/\mathrm{ZnO}$ system is an interesting and
promising material combination for the realization of multi-functional FM/SC
heterostructures.

This work is supported by the DFG via the priority programs 1157 and 1285
(project Nos. GR~1132/13 \& 14), and GO 944/3-1. We also acknowledged support
of the Excellence Cluster "Nanosystems Initiative Munich (NIM)".


\begin{thebibliography}{99}
\footnotesize
\bibitem{Wolf:2001a} S.A. Wolf, D. D. Awschalom, R. A. Buhrman, J. M.
    Daughton, S. von Molnar, M. L. Roukes, A. Y. Chtchelkanova, and D. M. Treger,
    Science {\bf 294}, 1488 - 1495 (2001).

\bibitem{Zutic:2004a}
    I. Zutic, J. Fabian, S. Das Sarma,
    Rev. Mod. Phys. {\bf 76}, 323-410 (2004).

\bibitem{Schmidt:2000} G. Schmidt, D. Ferrand, L. W. Molenkamp, A. T. Filip,
    and B. J. van Wees, Phys. Rev. B {\bf 62}, R4790 (2000); see also
    J. Phys. D: Appl. Phys. \textbf{38}, R107-R122 (2005)

\bibitem{Hanbicki:2003} A. T. Hanbicki, O. M. J. van´t Erve, R. Magno, G.
    Kioseoglou, C. H. Li, B. T. Jonker, G. Itskos, R. Mallory, M. Yasar, and A.
    Petrou, Appl. Phys. Lett. {\bf 82}, 4092 (2003).

\bibitem{Zhang:1991} Z. Zhang, and S. Satpathy, Phys.~Rev.~B {\bf 44}, 13319
    (1991)

\bibitem{Dedkov:2002} Yu. S. Dedkov, U. R\"{u}diger, and G. G\"{u}ntherodt, Phys.
    Rev. B {\bf 65}, 064417 (2002).

\bibitem{Fonin:2005} M. Fonin, R. Pentcheva, Yu. S. Dedkov, M. Sperlich,
    D. V. Vyalikh, M. Scheffler, U. R\"{u}diger, and G. G\"{u}ntherodt, Phys. Rev.
    B {\bf 72}, 104436 (2005).


\bibitem{Reisinger:2004a} D. Reisinger, P. Majewski, M. Opel, L. Alff,
    R. Gross,
    Appl. Phys. Lett. {\bf 85}, 4980 (2004).

\bibitem{Kennedy:1999} R. J. Kennedy and P. A. Stampe, J. Magn. Magn. Mater.
    195 \textbf{284} (1999); see also J. Phys. D: Appl. Phys.
    {\bf 32}, 16 (1999).

\bibitem{Reisinger:2003b} D. Reisinger,  M. Schonecke, T. Brenninger, M.
    Opel, A. Erb, L. Alff, R. Gross,
    J. Appl. Phys. {\bf 94} 1857 (2003).


\bibitem{Lu:2004} Y. X. Lu, J. S. Claydon, and Y. B. Xu, S. M. Thompson, K.
    Wilson, G. van der Laan, Phys.~Rev.~B {\bf 70}, 233304 (2004); see also
    J. Appl. Phys. {\bf 95}, 7228 (2004).

\bibitem{Lu:2005} Y. X. Lu, J. S. Claydon, E. Ahmad, Y. Xu, S. M. Thompson, K.
    Wilson, and G. van der Laan, IEEE Trans. on Magn., \textbf{41}, 2808
    (2005); see also J. Appl. Phys. {\bf 97}, 10C313 (2005).

\bibitem{Watts:2004} S. M. Watts, K. Nakajima, S. van Dijken, and J. M. D.
    Coey, J. Appl. Phys. \textbf{95}, 7465 (2004); see also Appl. Phys. Lett. \textbf{86}, 212108 (2005).


\bibitem{Boothman:2007} C. Boothman, A. M. S\'{a}nchez, S. van Dijken, J. Appl.
    Phys. {\bf 101}, 123903 (2007).

\bibitem{Dellile:1996} F. Dellile, B. Diny, J.-B. Moussy, M.-J. Guittet, S.
    Gota, M.Gautier-Soyer, C. Marin, J. Magn. Magn. Mater. {\bf 294}, 27
    (2005).

\bibitem{Margulies:1996} D. T. Margulies, F. T. Parker, and F. E. Spada, R. S.
    Goldman, J. Li and R. Sinclair, A. E. Berkowitz, Phys.~Rev.~B {\bf 53},
    9175 (1996).

\bibitem{Ogale:1998} S. B. Ogale, K. Ghosh, R. P. Sharma, R. L. Greene, R.
    Ramesh, and T. Venkatesan, Phys. Rev. B {\bf 57}, 7823 (1998).


\bibitem{Yamaguchi:2002} I. Yamaguchi, T. Terayama, T. Manabe, T. Tsuchiya, M.
    Sohma, T. Kumagai, and S. Mizuta, J. Solid State Chem. {\bf 163}, 239
    (2002).

\bibitem{Farrow:2003} R. F. C. Farrow, P. M. Rice, M. F. Toney, R. F. Marks, J.
    A. Hedstrom, R. Stephenson, M. J. Carey, and A. J. Kellock, J. Appl. Phys.
    {\bf 93}, 5626 (2003)

\bibitem{Moussy:2004} J.-B. Moussy, S. Gota, A.Bataille, M.-J. Guittet, and M.
    Gautier-Soyer, F. Dellile and B. Dieny, F. Ott and T. D. Doan, P. Warin and
    P. Bayle-Guillemoud, C. Gatel and E. Snoeck, Phys. Rev. B {\bf 70}, 174448
    (2004).

\bibitem{Klein:1999b} J. Klein, C. H\"{o}fener, L. Alff, and R. Gross, Supercond.
    Sci. Technol. \textbf{12}, 1023 (1999).

\bibitem{Fleet:1981} M.E. Fleet, Acta Cryst. B \textbf{37}, 917 (1981).


\bibitem{Reeber:1970}  R. R. Reber, J. Appl. Phys. {\bf 41}, 5063 (1970).


\bibitem{Kirfel:1990} A. Kirfel, K. Eichhorn. Acta Crystallogr. A
    {\bf 46}, 271-283 (1990).



\end{thebibliography}
\end{document}